\documentclass[12pt]{article}

\usepackage[utf8]{inputenc}

\usepackage[english, russian]{babel}
\usepackage{braket}
\usepackage{amsmath}
\usepackage{mathtools}
\usepackage{breqn}
\usepackage{graphicx}
\usepackage{amsfonts}
\usepackage{indentfirst}
\usepackage{amssymb}

\renewcommand{\d}{\mbox{d}}
\makeatletter \@addtoreset{equation}{section} \makeatother

\newcommand{\un}{{\underline{n} }}
\newcommand{\um}{{\underline{m} }}

\newcommand{\be}{\begin{equation}}
\newcommand{\ee}{\end{equation}}
\newcommand{\bee}{\begin{eqnarray}}
\newcommand{\beee}{\begin{array}}
\newcommand{\eee}{\end{eqnarray}}
\newcommand{\eeee}{\end{array}}

\newcommand{\ie}{{\it i.e.,} }

\begin{document}
	
	\begin{flushright}
		%\texttt{Started on May 14}
		%\\
		{\small FIAN/TD/16-2021}
	\end{flushright}
	\vspace{1.7 cm}
	
	\begin{center}
		{\large\bf On the variational principle in the unfolded dynamics}
		
		\vspace{1 cm}
		
		{\bf A.A. Tarusov${}^{1,2}$ and   M.A.~Vasiliev${}^{1,2}$}\\
		\vspace{0.5 cm}
		{\it
			${}^1$ I.E. Tamm Department of Theoretical Physics, Lebedev Physical Institute,\\
			Leninsky prospect 53, 119991, Moscow, Russia}
		
		\vspace{0.7 cm} \textit{ ${}^2$
			Moscow Institute of Physics and Technology, Institutsky pereulok 9, 141701, Dolgoprudny, Moscow region, Russia}
		\vspace{0.6 cm}
		% vasiliev@lpi.ru \\
	\end{center}

	\selectlanguage{english}
	
	\begin{abstract}
The interplay between off-shell and on-shell unfolded systems is analyzed.
		The formulation of invariant constraints that put an off-shell system on shell
is developed by adding new variables and derivation  in the target space, that extends the
original $Q$-derivation of the unfolded
system to a bicomplex. The analogue of the Euler-Lagrange equations
in the unfolded dynamics  is suggested.
 The general class of  invariant on-shell equation constraints
is defined in cohomological terms. The necessary and sufficient condition for
the on-shell equation constraints being Euler-Lagrange for some Lagrangian system is proven.
The proposed construction is illustrated by the scalar field example.
	\end{abstract}

	\newpage
	
	\tableofcontents
	
	\newpage
	
	\section{Introduction}
	
	Unfolded formalism provides a useful tool for the description of
  physical systems and, in particular, higher-spin gauge theories (see \cite{4authors}
  for a review). Unfolded equations are formulated solely in terms of differential
  forms $W^\Lambda(x)$ in the world-sheet space $M,$ which guarantees their coordinate independence.
  In application to field theories, unfolded systems involve
  infinite sets of first-order partial differential equations demanded to obey
   the compatibility conditions.
  Furthermore, symmetries of the physical system act geometrically on the differential forms of the unfolded system. The maximally symmetric background geometry is described by the Maurer-Cartan equations representing the simplest  unfolded equations \cite{Vasiliev:2005zu}.
	
	An off-shell unfolded system is a set of conditions expressing  higher-rank tensor components
in the system via  higher-order derivatives of the
{\it ground fields}, that are not expressed via derivatives of the other fields, in a way respecting the compatibility conditions
and gauge symmetries at every step.  (For recent development of the off-shell higher-spin  systems see e.g. \cite{Misuna:2019ijn,Misuna:2020fck}.)
The unfolded off-shell system itself describes the gauge
 symmetries of the system, but no dynamics,
\ie partial differential equations on the ground fields. Non-trivial dynamics  in the on-shell system results from imposing additional  constraints on the fields $W^\Lambda(x)$, that effectively restrict some of the  derivatives of the ground fields of the off-shell system.
	
Unfolded equations can be naturally reformulated in terms of a $Q$-derivation that
associates invariant functionals with the $Q$-cohomology \cite{Vasiliev:2005zu}.
In the off-shell case, the latter  represent various gauge invariant actions.

	This approach allows one to reconstruct the already known field-theoretic examples and has been successfully used for the description of the higher-spin gauge theory \cite{Vasiliev:1988xc,Vasiliev:1988sa,Vasiliev:1992av}. However, the algorithm identifying
 the set of conditions that put the system on-shell was not available directly within the
 unfolded formalism, except for the case when the
action and associated Batalin-Vilkovisky formalism were explicitly known allowing the authors
of \cite{Barnich:2004cr,Barnich:2010sw,Grigoriev:2012xg} to reconstruct the associated unfolded systems.
  The aim of this paper is in a certain sense opposite. We start with an off-shell unfolded system
 and analyze a possible form of the unfolded  Euler-Lagrange equations  to
  establish a relation between the off-shell unfolded systems and the related on-shell
systems associated with the action principle. More precisely, our aim is to represent
in the unfolded form the constraints themselves that put the system on shell.  Let us stress that, starting from a given  off-shell unfolded system, we address the question how to introduce
constraints on the fields that effectively impose the partial differential equations on the ground fields. The latter may or
may not be Euler-Lagrange for some action for the ground fields. This has to be contrasted with the papers \cite{Vasiliev:1988sa,Boulanger:2011dd}, where the unfolded system itself is derived from the action principle. It should be emphasized that our approach is applicable to any unfolded systems that may have different  form from those
of \cite{Barnich:2004cr,Barnich:2010sw,Grigoriev:2012xg}, such as those using spinor variables like in \cite{Vasiliev:1988xc,Vasiliev:1988sa,Vasiliev:1992av,Misuna:2019ijn,Misuna:2020fck}.

To this end we introduce additional variables $\delta W^\Lambda$, identified with the differentials of the fields $W^\Lambda$ in the target space $\mathcal{M}$ where $W^\Lambda$ are local coordinates. Extending the derivation $Q$ that
determines the form of the original unfolded equations to $\mathcal {Q}$, acting on both $W$ and $\delta W$, and
introducing the de Rham derivation $\delta$ that acts in $\Omega(\mathcal{M})$ we find that they form
a bicomplex
$$
\mathcal{Q}^2=0\,,\qquad \delta^2 =0 \,,\qquad \mathcal{Q} \delta + \delta \mathcal{Q} =0\,.
$$
In these terms we define the concept of invariant on-shell constraints that put the
unfolded system
on shell and find the necessary and sufficient conditions at which the resulting
system is  Lagrangian  in terms of the ground fields.

Though the general features of our formalism are
somewhat reminiscent of  the variational bicomplex described in \cite{Anderson1992},
 the details are different and the explicit relation is an interesting question for
the  future study.

	The paper is structured as follows:
	In Section 2 a brief recollection of the unfolded dynamics is provided with some emphasis on  the invariant functionals. In Section 3 variations of fields are introduced and the unfolded system is extended to incorporate them. In Section 4 the variational operator is added, that is shown to form a bicomplex with the extended homological vector field $\mathcal{Q}$ of the original unfolded system. In Section 5 the inverse problem of existence of a Lagrangian for given equations of motion is solved. In Section 6 it is shown that the suggested formalism allows for a natural inclusion of background fields. Section 7 illustrates the proposed construction by a toy example of a scalar field. Section 8 contains brief conclusions.
	
	\section{Unfolding} \label{Unfolding}	The possibility of decreasing the order of a differential equation by introducing new variables is well known. This approach can be generalized to partial differential equations by using the jet formalism \cite{Vinogradov}. Unfolded formalism is a
	further extension of this idea appropriate for the gauge theories in the framework of
	gravity.
	
	A system of unfolded equations \cite{Vasiliev:1988xc,Vasiliev:1988sa}
	(see also \cite{Boulanger:2008up}) is a generalization of the  first-order
	formulation of a system of ordinary
	differential equations, achieved by replacing the partial derivative by the de Rham derivative. This implies that the dynamical variables are now space-time differential forms $W^\Omega(x)$, which allows us to rewrite the system of equations in the form
	\begin{equation}
		\label{dw}
		\d W^\Omega(\xi, x) = G^\Omega(W(\xi, x))\,,\qquad \d:= \xi^\un \frac{\partial}{\partial x^\un}\,,
	\end{equation}
	where $\xi^\un$ is the differential, used as a placeholder for $dx^\un $, and $G^\Omega(W(\xi, x))$ is  some function of
	$W^\Omega(\xi,x)$ containing exterior products of the differential forms $W^\Omega(\xi, x)$ at the same $x$
	(no space-time derivatives in $G^\Omega(W)$; wedge product symbols are implicit while the variables
$\xi^\un $ are anticommuting $\xi^\un\xi^\um = -\xi^\um \xi^\un)$. The
     functions $G^\Omega(W(\xi, x))$ cannot be  arbitrary,
	as the de Rham derivation is nilpotent and thus the compatibility condition $\d G^\Omega=0$ must hold. This demands
	\begin{equation}
		\label{con}
		G^\Lambda( W) \frac{\partial G^\Omega(W)}{\partial W^\Lambda} = 0\,.
	\end{equation}
	This constraint allows one to show that
	system (\ref{dw}) is invariant under
	the following gauge transformations
	\begin{equation}
	\delta_{gauge}  W^\Omega(x) = \d\epsilon^\Omega(x) + \epsilon^\Lambda(x) \frac{\partial G^\Omega(W(x))}{\partial W^\Lambda (x)}\,, \label{gaugetransform}
	\end{equation}
	where
	$$
	{\rm deg}\,\epsilon^\Lambda = {\rm deg}\, W^\Lambda-1
	$$
	with ${\rm deg}\, \omega$ being the degree of the differential form $\omega$.
	
	Generally speaking, gauge invariance only takes place in the so-called universal systems, in which the compatibility conditions hold as a consequence of the system itself, without taking into account the number of space-time dimensions, \ie not using the fact that $(d+1)$-forms vanish in $d$-dimensional space. Indeed, for systems that are not universal, partial derivative $\frac{\partial G^\Omega}{\partial W^\Lambda}$ might not have any sense, leading to a non-zero derivative of a vanishing term, represented by a $(d+1)$-form. We shall assume that the systems considered in this paper are, indeed, universal as is the case for higher-spin theories.
	
	For universal systems it is  possible to introduce a $Q$-derivation
	(homological vector field) of the  form \cite{Vasiliev:2005zu}
	\begin{equation}
		Q = G^\Omega(W) \frac{\partial}{\partial W^\Omega}\,,
	\end{equation}
	which turns out to be nilpotent,
	$$
	Q^2 =0\,
	$$
	as a consequence of the compatibility conditions (\ref{con}). In these terms, any universal unfolded system can be rewritten as
	\be\label{QF}
\d F(W(x))= QF(W(x))\,,
\ee
	where $F(W)$ is an arbitrary function of the differential forms in question.

	This form of the unfolded equations implies that they equate the de Rham
derivation on the world-sheet space $M$ with coordinates $x^\un$  and the derivation $Q$  on the target space  $\mathcal{M}$ with local coordinates $W^\Lambda$.

Note that the form of the equations (\ref{dw}) implies that $G^\Lambda $ must have degree $p+1$
if $W^\Lambda$ had degree $p$. Hence,
\be
\label{deg}
deg\, Q =1\,.
\ee
	
	Consider, following  \cite{Vasiliev:2005zu}, a $Q$-closed Lagrangian function $\mathcal{L}$,
\be
\label{QL}
Q \mathcal{L}(W)=0\,
\ee
along with a $(d-1)$-form $E$ and a $d$-form $L$, such that
		\begin{eqnarray}
&&	\d E = L - \mathcal{L}(W),\\
&&		\d L = 0\,.
	\end{eqnarray}
	The action defined as the integral of $L$ over a $d$-cycle $M^d$
		\begin{equation}
		S = \int_{M^d} L\,
	\end{equation}
is gauge invariant under the gauge transformations associated via (\ref{gaugetransform})
with the fields $E$ and $L$ with the gauge parameters $\epsilon_E$ and $\epsilon_L$,
respectively,
 as well as under the gauge transformation (\ref{gaugetransform})  with the gauge parameters
 $\epsilon^\Lambda(x)$ associated with $W^\Lambda(x)$\,,
		\begin{eqnarray}
		&& \delta_{gauge} E = \d \epsilon_E +  \epsilon_L - \epsilon^\Lambda \frac{\partial \mathcal{L}}{\partial W^\Lambda}, \\
		&&\delta_{gauge} L = \d \epsilon_L\,.
	\end{eqnarray}
The gauge transformation with the parameter $\epsilon_L(x)$
 allows us to gauge fix $E$ to zero, giving
		\begin{equation}
		S = \int_{M^d} \mathcal{L}\,.
	\end{equation}
The conclusion is that invariant functionals for any unfolded
system are associated with the $Q$-cohomology \cite{Vasiliev:2005zu}. Indeed, $Q$-exact
Lagrangians describe total derivatives by virtue of 	(\ref{QF}).

	One distinguishes between  off-shell  and  on-shell unfolded systems. In the on-shell case, the system contains dynamical equations of motion (PDE) on certain ground (primary) fields. An off-shell
system just expresses  higher-rank tensor components
in the system via  higher-order derivatives of the
ground fields along with the compatibility condition for the latter expressions. It should be stressed that the same universal unfolded equations considered in the world-sheet manifolds $M^d$ of different dimensions may have different
interpretations being on-shell in, say, $M^{d'}$ and off-shell in  $M^{d''}$ with $d'> d''$ \cite{Vasiliev:2012vf}.

	Unfolded formalism allows for a natural  description of background geometry by Maurer–Cartan equations having the form of the simplest unfolded equations. Indeed, let $g$  be a Lie algebra with a basis \{$T_a$\} and Lie bracket $[\,,]$. Assuming  $W = w = w^a T_a$ and $G= -  \frac{1}{2} [w\,,w]$, one can check that equation (\ref{dw}) takes the form of the flatness (zero-curvature) condition
		\begin{equation}\label{MC}
		\d w + \frac{1}{2} [w\,,w] = 0\,,
	\end{equation}
which is just the Maurer–Cartan equation, the solutions to which are known to
represent $g$-invariant backgrounds in the Cartan gravity terms. (For more
detail see e.g. \cite{4authors} and references therein.)

	\section{Unfolded variations}

Consider an off-shell unfolded system. To achieve non-trivial dynamics by
restricting some of the lower-order derivatives, some extra conditions have to be imposed on the fields in the system. Let us illustrate how this works in a Lagrangian system. Let the Lagrangian  be represented by a  form $\mathcal{L}$ obeying the $Q$-closure  condition (\ref{QL}). As usual, the dynamics
results from extremizing the action functional $S = \int \mathcal{L}$, \ie
imposing the Euler-Lagrange equations
\begin{equation}
	\delta S (W)= \int_{M^d} \delta W^\Lambda(x) \frac{\partial \mathcal{L} (W)}{\partial W^\Lambda(x)} = 0\,,
\end{equation}
where $\delta W^\Lambda(x)$ is the variation  of $W^\Lambda(x)$
\begin{equation}
\label{var}
	\delta f(W) = \delta W^\Lambda(x) \frac{\partial f(W)}{\partial W^\Lambda(x)} .
\end{equation}

For the future convenience we will assume that $\delta W^\Lambda$ has grading shifted by
one compared to $ W^\Lambda$, which
will allow us to interpret the variations $\delta W^\Lambda$ as one-form differentials in the
target space with local coordinates $W^\Lambda$. More in detail,
in addition to the grading $deg$ identified with the degree $p$ of a
world-sheet differential form $W^\Lambda$,
\be
deg (W^\Lambda(x)) = p^\Lambda = deg (\delta W^\Lambda(x))\,,
\ee
we introduce the target space differential form grading
\be
Deg\, W^\Lambda(x)=0\,,\qquad Deg \,(\delta W^\Lambda(x)) =1\,.
\ee

The  commutation properties are determined by the total grading
\begin{equation}
	\pi = Deg + deg\,, \qquad fg = (-1)^{\pi(f) \pi(g)} gf \,.
\end{equation}

Since the fields $W^\Lambda$ are not independent, being related by the off-shell  unfolded equations (\ref{dw}),
their variations have to obey the conditions resulting from the variation of (\ref{dw})
\begin{equation}
\label{ddq}
\d \delta W^\Lambda(x) = - \delta W^\Omega(x) \frac{\partial G^\Lambda(W)}{\partial W^\Omega(x)} \,,
\end{equation}
where the extra sign results
from the extra grading of $\delta W^\Lambda$ compared to $W^\Lambda$.

Clearly, since the equation (\ref{ddq}) is a consequence of the consistent equation
(\ref{dw}), the system of equations (\ref{dw}) and (\ref{ddq})  is consistent as well.
The idea is to unify the fields $W^\Lambda$ and their variations $\delta W^\Omega$ into an extended
set of fields
$$
\tilde W^A(x)=(W^\Lambda(x),\delta W^\Omega(x)) \,,
$$
 obeying the system of unfolded equations (\ref{dw}), (\ref{ddq}).

Let us introduce the operator $Q'$ for the right hand side of the equations (\ref{ddq}) (from here onwards we shall be omitting the $x$ dependencies if it does not create ambiguity)

\begin{equation}
Q':= - \delta W^\Omega \frac{\partial G^\Lambda(W)}{\partial W^\Omega}  \frac{\partial}{\partial\delta W^\Lambda}\,.
\end{equation}
 Since $Q$ and $Q'$ act only on their respective variables, they can be combined to rewrite equations in a uniform way
\begin{equation}
\label{dt}
	\d \tilde{W}^A(x) = \mathcal{Q} \tilde{W}^A(x) \,,\qquad \mathcal{Q}:= Q+Q' \,.
\end{equation}

It is not difficult to see that $\mathcal{Q} $ is also nilpotent as a result of the compatibility conditions for the equations (\ref{dw}),
\begin{equation}
	\mathcal{Q}^2 = 0 \,. \label{mathcalq^2}
\end{equation}
This simple fact plays the key role in our construction.

Note that, since $Q'$ acts trivially on the $\delta W^\Lambda$--independent functions, cohomologies of $Q$ are also cohomologies  of $\mathcal{Q}$
\be
H(Q)\subset H(\mathcal{Q})\,.
\ee

Clearly,
\be
deg \, \mathcal{Q} = 1\,,\qquad Deg\, \mathcal{Q} =0\,.
\ee

Since  $\delta W^\Lambda$ are treated as new fields in the unfolded system they have their own gauge parameters $\delta\epsilon$ and a gauge transformation law
\begin{equation}
	\delta_{gauge} \delta W^{\Omega} = \d (\delta \epsilon^\Omega) - (\delta \epsilon^\Lambda)  \frac{\partial G^\Omega(W)}{\partial W^{\Lambda}} - (-1)^{p^\Lambda}\epsilon^\Lambda \delta W^{\Phi} \frac{\partial^2 G^\Omega(W)}{\partial W^{\Phi}\partial W^{\Lambda} } \,.
\end{equation}
The gauge transformations (\ref{gaugetransform}) for the fields $W^\Omega$  remain unchanged.

\section{Target space bicomplex}

Let us now introduce another nilpotent operator
\begin{equation}
\label{del}
 \delta := \delta W^\Lambda \frac{\partial}{\partial W^\Lambda}\,,\qquad \delta^2=0\,,
\end{equation}
which is the de Rham derivation in the target space $\mathcal{M}$
of  $W^\Lambda$, that acts on $\Omega(\mathcal{M})$ and has gradings
\be
\label{grd}
deg\, \delta =0\,,\qquad Deg \,\delta =1\,.
\ee

A remarkable fact is that
$\mathcal{Q}$ and $\delta$ anticommute:
\begin{equation}
\label{tQd}
\left\{ \mathcal{Q}\,,\delta \right\}=0\,,
\end{equation}
as one can see by an elementary straightforward computation
\begin{multline}
	 \left\{ G^\Lambda(W) \frac{\partial}{\partial W^\Lambda} - \delta W^\Omega \frac{\partial G^\Lambda(W)}{\partial W^\Omega} \frac{\partial}{\partial \delta W^\Lambda} , \delta W^\Phi \frac{\partial}{\partial W^\Phi} \right\} = - \delta W^\Omega \frac{\partial G^\Lambda(W)}{\partial W^\Omega} \frac{\partial}{\partial W^\Lambda} + \\
	 + \delta W^\Phi \frac{\partial G^\Lambda(W)}{\partial W^\Phi} \frac{\partial}{\partial W^\Lambda}
	 - \Big(\delta W^\Phi \frac{\partial}{\partial W^\Phi}  \delta W^\Omega \frac{\partial G^\Lambda(W)}{\partial W^\Omega}\Big) \frac{\partial}{\partial \delta W^\Lambda} = 0 \,, \label{anticommutatorcalculation}
\end{multline}
where all terms with two differentiations on the {\it r.h.s.} are dropped since they vanish
in the anticommutator of any odd vector fields.

This result is not too surprising however, as $\mathcal{Q}$ can be seen as the Lie derivation along a vector field $G^\Lambda$ via the Cartan formula:
	\begin{equation}
		\mathcal{Q} := \left[i_G, \delta\right] = \left[ G^\Omega \frac{\partial}{\partial \delta W^\Omega}, \delta W^\Lambda \frac{\partial}{\partial W^\Lambda}\right]\,,
	\end{equation}
which makes (\ref{tQd}) obvious by virtue of the second relation in (\ref{del}).

{Equation (\ref{tQd})} implies that the total derivation
\be
Q_{tot}:=\mathcal{Q} + \delta
\ee
is also nilpotent
\be
Q_{tot}^2=0\,.
\ee
The derivations $\mathcal{Q}$ and $\delta$ form a bicomplex.

According to (\ref{var}) $\delta$ is nothing else but the variation operator.
As a consequence of (\ref{tQd}) for every $Q$-closed $\mathcal{L}$, $\delta \mathcal{L}$ is $\mathcal{Q}$-closed. Hence for every invariant functional of the original unfolded system, its variation is $\mathcal{Q}$-closed.
In these terms the Euler-Lagrange equations demand $\mathcal{L}$ to be $\delta$-closed, up to
a total derivative in the world-sheet space $M^d$. Since, by virtue of the unfolded equations (\ref{dt}), space-time derivative equates to the action of $\mathcal{Q}$, it implies  that
\begin{equation}\label{EL}
  \delta \mathcal{L}+\mathcal{Q} \Phi =0\,,
\end{equation}
where $\Phi$ is adjusted in such a way that the {\it l.h.s.} of (\ref{EL}) only contains
the variation of the ground fields among $W^\Lambda$.
Recall that, similarly to the fact that only
a few ground fields among $W^\Lambda$ can be treated as independent  while all others are expressed via their space-time derivatives by virtue of (\ref{dw}), only the variations of the
ground fields are independent, while all others are their descendants to be eliminated  by
an appropriate $\mathcal{Q} \Phi$. This mechanism is illustrated by the
scalar field example in Section \ref{scal}.

\section{Inverse problem}
\label{inverse}

Let $\mathcal{E}(\delta W, W)$ be a one-form in the target space $\mathcal{M}$ and $d$-form in $M^d$
\be
Deg\,\mathcal{E} =1\,,\qquad deg \mathcal{E} = d\,.
\ee
Let us address the question whether the condition
\be
\label{econ}
\int_{M^d} \mathcal{E}(\delta W(x), W(x))=0 \,
\ee
with arbitrary $\delta W(x)$ and $ W(x)$, subject to their unfolded equations (\ref{ddq})
and (\ref{dw}), respectively, can describe some gauge invariant
equations. Analogously to the case of actions considered in \cite{Vasiliev:2005zu},
the integral (\ref{econ}) is gauge invariant iff $\mathcal{E}$ is $\mathcal{Q}$-closed
\be
\label{QE}
\mathcal{Q} \mathcal{E}(\delta W, W) =0\,.
\ee
Note that Euler-Lagrange equations (\ref{EL}) fulfill this condition.
$\mathcal{Q}$-exact  $\mathcal{E}(\delta W, W)$ trivializes upon integration in (\ref{econ}) due to (\ref{dt}).
Therefore, nontrivial gauge invariant equations are represented by
$Deg =1$ $H(\mathcal{Q})$.

By (\ref{EL}) the necessary condition for $\mathcal{E}$ to represent Euler-Lagrange equations therefore
is
\be
\label{consi}
\delta \mathcal{E} =  \mathcal{Q} \delta \Phi
\ee
with some $\Phi$. If
$\delta \mathcal{E} \neq  \mathcal{Q} \delta \Phi $, the equations (\ref{econ}) associated with
 $\mathcal{E}$ are not Lagrangian.

 If equation (\ref{consi}) holds true, $\mathcal{E}$ can be shown to
 represent Lagrangian equations. Indeed, using the Poincar\'e lemma for the target space
 de Rham derivation $\delta$, (\ref{consi}) can be rewritten in the form
\be
\label{consi1}
 \mathcal{E} +  \mathcal{Q}  \Phi = \delta \chi\,.
\ee
The fact that $\mathcal{E}$ is $\mathcal{Q} $-closed yields
\be
\label{chi}
\delta \mathcal{Q} \chi=0\,.
\ee
Since $\chi$ and hence $\mathcal{Q}\chi$ are
 target-space zero-forms, it follows from (\ref{chi}) that $\mathcal{Q}\chi$ is $W$ independent. However, for a non-zero $\mathcal{Q} \chi$, $deg (\mathcal{Q} \chi)>0$ which is not possible for $\mathcal{Q} \chi$ being a constant. From this  it follows that $\mathcal{Q} \chi$ is indeed zero, {\it i.e.},
\be
\chi = \mathcal{L} +\mathcal{Q} \phi\,
\ee
with some $\mathcal{L}$ representing $H(\mathcal{Q})$. This gives
\be
\mathcal{E}=\delta \mathcal{L} +\mathcal{Q} (\Phi+\delta\phi)\,,
\ee
which is just the variation (\ref{EL}) with the
appropriately redefined $\Phi$. Hence we arrive at the following\\
{\it Theorem:}\\
 {The equations
(\ref{econ}) associated with $\mathcal{E}$ are  Euler-Lagrange iff the condition (\ref{consi}) holds true.}\\
This is one of the central results of the paper.

After being reduced to the form
\be
\label{eq}
\int_{M^d} \delta W^i_{gr}(x) e_i(W)=0\,,
\ee
with the help of (\ref{ddq}) and addition of total derivatives in the form
of $\mathcal{Q}$-exact terms (\ie expressing all $\delta W^\Lambda$ via derivatives of
$\delta W^i_{gr}$ by virtue of (\ref{ddq})),
equation (\ref{econ}) yields constraints on the fields $W^\Lambda$ and hence $W^i_{gr}(x)$,
\be
\label{e}
e_i (W)=0,
\ee
with some $e_i (W)$. These are usual Euler-Lagrange equations.

  It should be stressed that the resulting equations
(\ref{e}) are by construction invariant under the gauge transformations of the
original off-shell system which leave invariant the  on-shell
submanifold   (\ref{e}) of $\mathcal{M}$. Among these gauge transformations
some may themselves be proportional to $e_i(W)$. These act trivially on the
fields on-shell analogously to on-shell-trivial gauge transformations that appear in the so-called
open algebras.

Finally, let us note that instead of imposing the on-shell conditions (\ref{econ}) one can
quotient them away dropping all terms proportional to such expressions. Though, naively,
the two schemes are equivalent, in the infinite-dimensional case they may be different. For instance, in the higher-spin theory of \cite{Vasiliev:2003ev} (see also \cite{4authors})
the quotient approach was applied for a certain ideal in the HS algebra, while the constraint approach was considered in  \cite{Sagnotti:2004pod}.

\section{Background fields}

So far it was assumed that all fields in the system are dynamical. This is
the case in theories involving gravity but not true in presence of background
fields that describe  background space. For instance,
to describe Minkowski space in a coordinate independent way
the background geometry is described
in the unfolded  Maurer-Cartan form
for the Poincar\'e algebra:
\begin{eqnarray}
\label{po}
	\d e^n - \omega^{nm} e_m=0\,,\qquad
	\d \omega^{nm} - \omega^{nk} \omega_k{}^m=0\,, \label{Mauerer}
\end{eqnarray}
where $e^n$ and $\omega^{mn}$ are the vierbein and Lorentz connection one-forms, respectively.
In Cartesian coordinate system
$e^m=dx^m $ and $\omega^{nm}=0$.

Generally, we distinguish between the background fields $w^\alpha(x)$ and
the dynamical fields $W^\Omega(x)$, assuming that
unfolded equations have the form
\begin{equation}
\label{w}
\d w^\alpha(x) = g^\alpha (w(x))\,,
\end{equation}
\begin{equation}
\label{W}
\d W^\Omega(x) = G^\Omega (w(x),W(x))\,.
\end{equation}
The system (\ref{w}) is autonomous,
\begin{equation}
\label{Gw}
\frac{\partial g^\alpha}{\partial W^\Omega} = 0
\end{equation}
 with the background fields
not  varied
\begin{equation}
\label{dw1}
\delta w^\alpha = 0\,.
\end{equation}

 The system for dynamical fields (\ref{W}) depends
on both background fields and dynamical fields. Different choices of
the background fields are usually equivalent. For instance, different
solutions to equations (\ref{po}) with non-degenerate frame $e^n$ describe
the same Minkowski space-time in different coordinate systems.

 The variation problem in presence of the background fields is set
 with respect to the dynamical fields with the variations $\delta W^\Omega$
 and no variations for $w^\alpha$.
 Other details are not affected by the presence of the background fields
 due to the triangular form
 of the equations (\ref{w}), (\ref{W}).
Namely, using the compatibility conditions
\begin{equation}
	g^\alpha(w) \frac{\partial g^\beta(w)}{\partial w^\alpha} = 0 \,, \label{backgroundcompat}
\end{equation}
\begin{equation}
	g^\alpha(w) \frac{\partial G^\Omega(w,W)}{\partial w^\alpha} + G^\Lambda(w,W) \frac{\partial G^\Omega(w,W)}{\partial W^\Lambda} = 0 \, \label{dynamiccompat}
\end{equation}
along with (\ref{dw1}) one can easily check that
\begin{equation}
	Q = g^\alpha(w) \frac{\partial}{\partial w^\alpha} + G^\Omega(w,W) \frac{\partial}{\partial W^\Omega}\,, \qquad Q^2 = 0\,,
\end{equation}
\begin{equation}
	\mathcal{Q} = g^\alpha(w) \frac{\partial}{\partial w^\alpha} + G^\Omega(w,W) \frac{\partial}{\partial W^\Omega} - \delta W^\Omega \frac{\partial G^\Lambda(w,W)}{\partial W^\Omega}  \frac{\partial}{\partial\delta W^\Lambda}\,, \qquad \mathcal{Q}^2 = 0 \,. \label{newmathcalq}
\end{equation}

Since background fields are not  varied, it is also easy to check that $\delta$  introduced in the standard way
\begin{equation}
	\delta = \delta W^\Omega \frac{\partial}{\partial W^\Omega}\,, \qquad \delta^2 = 0 \,
\end{equation}
respects the bicomplex condition (\ref{tQd}).

\section{Scalar field example}
\label{scal}

Let us now illustrate the developed technique by a specific example of a scalar field, for which the unfolded formalism is well known \cite{Shaynkman:2000ts, Vasiliev:2005zu}, leading to equations
\begin{eqnarray}
\label{0eq}
	\d C(y,x) = dx^m \frac{\partial}{\partial y^m}\, C(y,x)
\end{eqnarray}
with  the generating function $C(y,x)$
\begin{equation}\label{Cpr}
	C(y,x) = \sum_{n=0}^{\infty} \frac{1}{n!} C_{m_1 ... m_n}(x) y^{m_1} ... y^{m_n}\,.
\end{equation}

It is not hard to see that this system is off-shell \cite{Vasiliev:2005zu}. Here the
genuine scalar field
$C(x):= C(0,x)$ is the ground field while all other descendent components  in the generating function (\ref{Cpr}) are expressed via derivatives of $C(x)$ by virtue of  (\ref{0eq}).

%Dynamics is added to this system by requiring $C_{mm} = 0$.
%{\bf Ob'yasnyayte podrobno.Ne ponyatno - chto v deystvii tozge $C_{mm} = 0$?}
This  system can be put on-shell by demanding
\be\label{onsh}
C_m{}^m = 0\,
,\ee
which imposes the Klein-Gordon equation $\Box C(x)=0$ on the ground field.

With background fields obeying the Maurer-Cartan equations for the Poincar\'e
algebra (\ref{po}) equation (\ref{0eq}) can be rewritten in the following coordinate-independent form:
\begin{equation}
	\d C (y,x) = (\omega_{nm} y^n \frac{\partial}{\partial y_m} + e^m \frac{\partial}{\partial y^m}) C(y,x) \,. \label{equationC}
\end{equation}

The extension to the variations yields
\begin{equation}
	\d \delta C(y,x) = - \delta C_m(y,x) (\omega_{n}{}^m y^n + e^m)\,,
\end{equation}

where

\begin{equation}
	\delta C_m(y,x) = \delta \frac{\partial}{\partial y^m} C(y,x)\,.
\end{equation}

The form of the invariant action for the original off-shell system is also known
\cite{Vasiliev:2005zu}
\begin{equation}
\label{L}
	\mathcal{L} = e^{n_1} e^{n_2} ... e^{n_d} \epsilon_{n_1 ... n_d} l(C, C_n, C_{nm}, ...),
\end{equation}
where $l$ is an arbitrary Lorentz invariant function, that is demanded by the condition
that all $\omega^{nm}$-dependent  terms have to cancel upon the action of $Q$,
which by (\ref{Mauerer}),(\ref{equationC}) has the form
\begin{eqnarray}
&&Q C(y,x)= (\omega_{nm} y^n \frac{\partial}{\partial y_m} + e^m \frac{\partial}{\partial y^m}) C(y,x) \,,\\
&&Qe^n = \omega^{nm} e_m \,,\\
&&Q\omega^{nm} = \omega^{nk} \omega_k{}^m \,.
\end{eqnarray}
The terms with the frame field vanish upon the action of $Q$
since $\mathcal{L}$ already contains  $d$ frame-fields.
In usual terms, $l$ is an arbitrary
Lorentz invariant function built from the derivatives of the field $C(x)$.

Taking into account variations $\delta C(y,x)$, which have similar equations, we can also write down the $\mathcal{Q}$, which acts as $Q$ on $C$, $e^n$, $\omega^{nm}$ and

\begin{equation}
	\mathcal{Q} \delta C(y,x)= - \delta C_m(y,x) (\omega_{n}{}^{m} y^n + e^m ) \,.
\end{equation}

Now let us consider a familiar Lorentz invariant action for the scalar field:
$$
l=C_m C^m\,.
$$
 As is well known its variation  yields equations of motion upon integrating out a total derivative. The latter is equivalent to the addition of an appropriate
 $\mathcal{Q}$-exact term.

More in detail, after taking the variation (working for simplicity in Cartesian coordinates)
\begin{multline}
	(\delta C_m) e^{n_1} e^{n_2} ... e^{n_d} \epsilon_{n_1 ... n_d} C^m = d  (\delta C_m) e^{m} e^{n_2} ... e^{n_d} \epsilon_{n_1 ... n_d} C^{n_1} = \\
	  - \mathcal{Q}(d  (\delta C) e^{n_2} ... e^{n_d} \epsilon_{n_1 ... n_d} C^{n_1}) + d  (\delta C) e^{m} e^{n_2} ... e^{n_d} \epsilon_{n_1 ... n_d} C_{m}{}^{n_1} \,,
\end{multline} which is clearly  $\mathcal{Q}$-closed
($d$ is  space-time dimension).
Also, it is by construction  $\delta$-exact by virtue of anticommutativity of both $e^i$ and $\delta C$.

Discarding the $\mathcal{Q}$-exact term, the variation takes the form
\be
\mathcal{E}= (\delta C) e^{n_1} e^{n_2} ... e^{n_d} \epsilon_{n_1 ... n_d} C^m_{m},
\ee
imposing the Klein-Gordon equation in the form (\ref{onsh}).

It should be noted, however, that, as discussed in \cite{Vasiliev:2005zu}, generally,
the choice of the Lagrangian $l$ in (\ref{L})
was arbitrary. Lagrangians from the same $\mathcal{Q}$-cohomology class give rise to equivalent equations of motion, while those from different classes  result in different equations of motion, featuring, for instance, mass terms, non-trivial potential, higher-derivative
and/or interaction terms. The next simplest example, $l = C_k C^k + m^2 C^2$, can be easily checked to reproduce the equations of motion for the massive scalar, while $l=C_{kl} C^{kl}$ leads to the higher-derivative equation
$\Box^2 C(x)=0$.

To illustrate general results of Section \ref{inverse} consider the following equations of motion for two  scalar fields
\begin{equation}
\label{nl}
	\mathcal{E} = e^{n_1} e^{n_2} ... e^{n_d} \epsilon_{n_1, ... n_d} ((\delta C_1) C^m_{1  m} + (\delta C_2) C^m_{2  m} + \delta C_2 C_1) \,.
\end{equation}
So defined $\mathcal{E}$ is Poincar\'e invariant and, hence, $\mathcal{Q}$-closed.
Its variation is
\begin{equation}
\delta \mathcal{E} = e^{n_1} e^{n_2} ... e^{n_d} \epsilon_{n_1, ... n_d} ((\delta C^m_{1  m}) (\delta C_1)  + (\delta C^m_{2  m})(\delta C_2) + \delta C_1 \delta C_2) \,.
\end{equation}

Since both $\delta C_1$ and $\delta C_2$ are ground fields,
no $\Phi$ satisfying (\ref{consi}) exists, which means in accordance with Section \ref{inverse}
that the equations are non-Lagrangian, which is indeed true. Note that the replacement of
$\delta C_2 C_1$ by $\delta C_2 C_1+\delta C_1 C_2$ makes the equation  (\ref{nl}) Lagrangian.

\section{Conclusion}

In this paper,  an extended unfolded system is suggested
by introducing  extra variables corresponding to variations $\delta W^\Omega$ of the
original fields $ W^\Omega$
and interpreted as differentials in the target space where $ W^\Omega$  serve as
local coordinates.
This allows us to define a new $\mathcal{Q}$-derivation in the extended system, giving rise to an action and the equations of motion resulting from the action variation.
Remarkably, $\mathcal{Q}$ and the de Rham derivation $\delta$ in the target space form a bicomplex
in terms of which the structure of invariant field equations can be analyzed. In particular,
we found a necessary and sufficient condition for the equation constraints on the ground (dynamical)
fields to be Lagrangian, \ie resulting from the action principle.

It should be stressed that our approach is entirely formulated within the
general unfolded dynamics formalism with no reference to its specific realization
and/or external ingredients like
for instance those used in \cite{Barnich:2004cr,Barnich:2010sw,Grigoriev:2012xg}.
This allows us to address all kind of dynamical questions within unfolded dynamics.
On the other hand,
the variation fields $\delta W^\Lambda$ introduced in this paper are somewhat analogous to the antifields in the Batalin-Vilkovisky formalism \cite{Batalin:1977pb}. It would be  interesting to make the relation between the
two approaches more explicit. The same concerns the variational bicomplex of \cite{Anderson1992}.

There are many questions for the future study. In particular, it would be nice to find a way
for reconstruction of the form of Lagrangian from the Lagrangian field constraint rather than
just proving the existence theorem. Also the interpretation of the higher cohomologies of the operator $\mathcal{Q}$ and the ($\mathcal{Q}$, $\delta$) bicomplex need further analysis. These questions are particularly interesting in the context of higher-spin gauge theory.

\section*{Acknowlegement}
We are grateful to V. Didenko, K. Ushakov and, especially,
O. Gelfond and M. Grigoriev for useful discussions and comments.
Also we are grateful to the referee for the suggestions improving the presentation.
The authors benefited a lot from the thematic program
”Geometry for Higher Spin Gravity: Conformal Structures, PDEs, and Q-manifolds”
at the Erwin Schr\"odinger International Institute for Mathematics and Physics, Vienna,
Austria.
This research was supported by the
Russian Science Foundation grant 18-12-00507.

\end{document}